\documentclass[twocolumn,showpacs,amsmath,amssymb]{revtex4}

\usepackage{graphics}
\usepackage[dvips]{graphicx}

\begin{document}

\title{The influence of non-minimally coupled scalar fields on \\ the dynamics of interacting galaxies}

\author{R. F. Gabbasov,
M. A. Rodr\'{\i}guez-Meza,  
J.L. Cervantes-Cota, and J. Klapp}

\address{Depto. de F\'{\i}sica, Instituto Nacional de Investigaciones
Nucleares, Apdo. Postal 18-1027, M\'{e}xico D.F. 11801, M\'{e}xico. \\
}

\date{\today}


\begin{abstract}
We study bar formation in galactic disks as a consequence of the
collision of two spiral galaxies under the influence of a
potential which is obtained from the Newtonian limit of a
scalar--tensor theory of gravity. We found that dynamical effects
depend on parameters ($\alpha$, $\lambda$) of the theory. In
particular, we observe that the bar is shorter for weaker tidal
perturbations, which in turn corresponds to smaller values of
$\lambda$ used in our numerical experiments.
\end{abstract}

\maketitle

\section{Introduction}
 Observations of spiral galaxies indicate that the presence of a
 central structure called
a {\it bar} is a common feature\cite{Elmegreen83}.  The
instability of isolated stellar disks in galactic models leads to
bar formation and is characterized by Toomre's stability parameter
$Q$\cite{Toomre64}.  The models with $Q<1$ are subject to bar
formation. However, we are interested in dynamical effects of
non--isolated systems which are found in clusters of gala\-xies.
In this regard, it has been suggested that the observed bar in many
spirals is the result of the gravitational interaction between two
or more nearby galaxies. For instance, Nogushi\cite{Nogushi87} has
found that, during the collision of two galaxies and between the
first and the second closest approaches, the disk takes on
transient bar shape.  The gravitational interaction between the
two galaxies gives rise to perturbations in the orbits of the
stars that results in the formation of the bar.

 Bar formation in the simulations of stellar disks depends
 upon various simultaneous effects. In the case of collisions, these
 factors are\cite{Salo91}: rotation curve shape, disk-halo
 mass ratio, perturbation force and geometry.
 However, simulations can suffer from numerical effects such as low
 spatial and temporal resolution, too few particles representing
 the system and an approximate force model. These effects can be
drastic: for example, observations show  that bars have typically
a length scale close to the exponential
 length of the disk\cite{Elmegreen85}, while the bar's semi-major
 axis obtained from numerical models is two to four times
 longer\cite{DeSe98,AtMi02}.

Recent observational data measured in the Cosmic Microwave
Background at various angular scales, in Supernovae Ia, in the
2dF galactic survey, and baryon acoustic oscilations, 
suggest \cite{BrCeSa04,Percival2009} that the Universe is
composed of about $4.6\%$ baryons in the form of gas and stars,
$23.2\%$ dark matter (DM) and $72.2\%$ dark energy, which is a kind of
cosmological constant and is responsible for the today accelerated
expansion of the Universe.  In this way, galaxies are expected to
possess these dark components and, in accordance with rotation
curves of stars and gas around the centers of spirals, this might
be in the form of halos, and  must contribute to at least 3 to 10
times the mass of the visible matter of spirals.

Regarding to the nature of DM, we know that DM has to be
non--baryonic. This is because nucleosynthesis abundances of light
elements are only consistent with the above--mentioned baryonic
fraction, and this is not sufficient at all to account for
rotational velocities of spirals. This fact opens up new
possibilities for explaining the nature of DM. In this sense, in
papers \cite{RodriguezMeza:2009sh, Rodriguez2003,Rodriguez-Meza04, Rodriguez2005} 
a model was proposed in which
a scalar field (SF) couples non-minimally to gravity to produce
locally a modified Newtonian theory of gravity. It turns out that
the dynamics is now determined by the Poisson equation coupled to
a Klein--Gordon equation for the assumed scalar field in the
galaxy.
 Thus, the boson mass of the scalar field modifies the Newtonian
 law of attraction, and the dynamics of DM is different from its Newtonian counterpart.
  In this scalar--tensor theory potential--density pairs
for various halo density profiles were computed\cite{Rodriguez-Meza04, Rodriguez2005}.

In the present work we use the above-mentioned results to study
the collision process of two spirals, each of which possess a
disk, bulge and dark halo, in order to estimate the effects of the
modified gravity theory on the bar's length and the orbital decay
of galaxies. 
A numerical treecode was developed by one of us (MARM) in order to compute
the evolution of an $N$-body system interacting with the standard gravity
plus the interaction of the scalar field\cite{RodriguezMeza:2009sh,Ruslan2006,Rodriguez2007,
RodriguezMeza:2008gc, RodriguezMeza:2008vf}.
We first study dynamical effects on isolated galaxy
models for three different
 interaction scales ($\lambda$).
 We found no significant changes in the morphology of models, but we did 
 in the total potential energy.
 Then, we analyze the formation of a bar during a parabolic collision of two identical
 galaxies and compare
 the results obtained in the SF model with three scales of $\lambda$ and with a pure 
 Newtonian
 interaction.

\section{Scalar--tensor theory and its Newtonian limit}
A typical scalar--tensor theory is given by the following Lagrangian
\begin{equation}
{\cal L} = \frac{\sqrt{-g}}{16\pi} \left[ -\phi R + \frac{\omega(\phi)}{\phi}
(\partial \phi)^2 - V(\phi) \right] + {\cal L}_M(g_{\mu\nu}) \; ,
\end{equation}
from which we get the gravity and  SF equations.
 Here $g_{\mu\nu}$ is the metric,
 ${\cal L}_M(g_{\mu\nu})$ is the Lagrangian matter and $\omega(\phi)$ and
 $V(\phi)$ are arbitrary functions of the SF.
 According to the Newtonian approximation, gravity and SF are weak,
 and the velocities of the stars are non--relativistic. Then, we expect to have small
 deviations of the SF around the background defined here as
 $\langle \phi \rangle \equiv G_0^{-1}$.
  If we define the perturbation $\bar{\phi} \equiv \phi - \langle \phi \rangle$,
 then the Newtonian approximation gives the equations \cite{Rodriguez-Meza04,Helbig91}
\begin{eqnarray}
\label{rio_eq_psiphibar} \nabla^2 \psi = 4\pi G_0 \rho  \quad ,
\quad \nabla^2 \bar{\phi} - m^2 \bar{\phi} =
 - 8\pi \alpha G_0\rho \; ,
\end{eqnarray}
where $\psi=\frac{1}{2}(h_{00}+\bar{\phi})$.
 Here we define $\lambda = \hbar/mc$, the Compton wavelength of the effective
 mass $m$ of some elementary particle (boson) given through $\omega(\phi)$ and
 the potential $V(\phi)$, and $\alpha \equiv 1/(3+2\omega(\phi))$ is the amplitude
 of the perturbed SF, $\bar{\phi}$. The above formalism is valid
 for any potential that can be expanded in Taylor series around $\langle \phi
 \rangle$.
 In what follows we will use $\lambda$ instead of $m^{-1}$.
 This mass can have a range of values depending on particular
 particle physics models.

General solutions of Eqs. (\ref{rio_eq_psiphibar}) can be found in terms
 of the corresponding Green's functions
 and the new Newtonian potential is
\begin{eqnarray}
\Phi_N &\equiv&  \psi
 - \frac{1}{2} \bar{\phi}
 = -G_0\int d{\bf r}_s
 \frac{\rho({\bf r}_s)}{|{\bf r}-{\bf r}_s|}
\nonumber \\
&& -\alpha G_0\int d{\bf r}_s \frac{\rho({\bf r}_s)
 {\rm e}^{- |{\bf r}-{\bf r}_s|/\lambda}}
{| {\bf r}-{\bf r}_s|} + \mbox{B.C.}
\label{rio_eq_gralPsiN}
\end{eqnarray}

Solutions of these equations for point masses are
\begin{eqnarray}
\bar{\phi} &=& 2 \alpha u_\lambda \, ,  \\
\Phi_N &=& -u - \alpha u_\lambda \,  ,
\end{eqnarray}
where
\begin{eqnarray}
u &=& \sum_s \frac{G_0 m_s}{| {\bf r} - {\bf r}_s |} \, , \\
u_\lambda &=& \sum_s \frac{G_0 m_s}{| {\bf r} - {\bf r}_s |} {\rm
e}^{ -| {\bf r} - {\bf r}_s |/\lambda } \; ,
\end{eqnarray}
with $m_s$ being a source mass. The total gravitational force on a
particle of mass $m_i$ is
\begin{equation}
{\bf F} = -\nabla \Phi_N = m_i {\bf a} .
\end{equation}
The potential $u$ is the Newtonian part and
$u_\lambda$ is the SF modification  which is of Yukawa type.

\section{Initial conditions}
We use the Monte-Carlo procedure to construct a galaxy model with
a Newtonian potential.
 A fully self--consistent model in the context of the SF is in preparation.
 The initial conditions of the galaxies are constructed following the
 model described by Barnes\cite{Barnes96}. In this model, both
 the bulge and halo are non-rotating, spherically symmetric
 and with an isotropic Gaussian distribution of velocities characterized
 by the velocity dispersions $\sigma_b$ and $\sigma_h$, respectively.
 The units are such that the local ($r\ll \lambda$) gravitational
constant is $G = G_0(1+\alpha) = 1$, and the units of mass,
longitude and time are $M = 2.2\times 10^{11}$ M$_{\odot}$, $R =
40$ kpc and $T = 250$ Myrs, respectively.
 The bulge density profile is\cite{Hernquist90}
\begin{equation}
\rho_b(r)=\frac{M_b a_b}{2\pi} \frac{1}{r(r+a_b)^3} \quad ,
\end{equation}
and the halo density profile is a Dehnen's family member with
$\gamma=0$ \cite{Dehnen93}
\begin{equation}
\rho_h(r)=\frac{3M_h}{4\pi} \frac{a_h}{(r+a_h)^4} \quad .
\end{equation}
The disk density profile is exponential \cite{Freeman70}
\begin{equation}
\rho_d(r,z)=\frac{M_d}{4\pi a_d^2 z_0} e^{-r/a_d} \mbox{sech}^2
\left( \frac{z}{z_0} \right) \quad .
\end{equation}
Here $M_b=0.0625$, $M_d=0.1875$, and $M_h=1.0$ are the total
 mass of the bulge, disk, and halo, respectively. The
scale lengths of the bulge, halo, and disk are $a_b=0.04168$,
 $a_h=0.1$, and $a_d = 1/12$, respectively, and $z_0=0.007$ is
 the scale height of the disk. The mass distributions were truncated at a radius
containing $95\%$ of the total mass, since they extend to
infinity. The compound galaxy was sampled with $N=40960$ equal
mass particles.
 The velocity distribution of the disk is given by the Schwarzschild distribution
 function
with the velocity dispersions
 $\sigma_R = 2\sigma_z\propto e^{-a_dr}$, and $\sigma_z$ given by the equilibrium
 condition of an infinite gravitating sheet; $\sigma_\phi$ is
 calculated from the epicyclic approximation.
 The Toomre's parameter of initial disk is $Q\approx 1$,
 so we have the disk which is marginally stable for axisymmetric
 perturbations, but not, however, against strong non-axisymmetric ones.

Observations suggest that the majority of the interacting galaxies
 are located on nearly parabolic orbits.
 For all the collisions, disks were located in the plane
 of parabolic
 orbits, calculated from parameterized equations of the two-body problem,
 with a pericentric separation $p=0.4$, and the time to
 pericenter $t_{p}=3.0$.
 The direction of rotation of one of the disks (disk 1)
 was in the same direction as the corresponding
 orbital-angular-momentum, i.e., direct motion.
 The other disk (disk 2) was in retrograde motion.
 The two colliding galaxies are initially identical.

\section{Numerical method}
For the time evolution we use a Barnes tree
 code type\cite{BarnesHut86} modified to include
 the Newtonian contribution of the scalar fields as given by
 Eqs. (4)-(8) \cite{Ruslan2006}.
 The forces were computed with a
 tolerance parameter $\theta=0.75$, and including the monopole term only.
 For the gravitational potential we used the standard Plummer model
\begin{equation}
\Phi \propto -\frac{1}{\sqrt{r^2+\epsilon^2}} \quad .
\end{equation}
Here, $\epsilon$ is the softening parameter taken in our simulations
 to be $\epsilon=0.015$.
 The equations of motion were integrated using the second order
 leap--frog algorithm with a fixed time step $\Delta t=1/256$.
 With these parameters we obtain a good energy conservation ($<0.2\%$)
 and, also, good angular momentum conservation ($<0.5\%$) for
 all runs presented here.

To characterize quantitatively the bar amplitude, we consider the
 distortion parameter defined as\cite{Shibataothers03}
\begin{equation}
\eta = \sqrt{\eta_+^2+\eta_\times^2} \quad ,
\end{equation}
where
\begin{equation}
\eta_+=\frac{I_{xx}-I_{yy}}{I_{xx}+I_{yy}} \quad , \quad
\eta_\times=\frac{2I_{xy}}{I_{xx}+I_{yy}} \quad ,
\end{equation}
and
\begin{equation}
I_{ij}=\sum_{k=1}^N m_k x_k^i x_k^j \quad , \quad i, j=(x,y) \quad .
\end{equation}
The particles that are outside of the spatial region of the
 original disk can affect the parameter under study.
 For instance, if
 we calculate the distortion parameter using all the particles
 in the disk, we have in both cases similar evolution curves.
 Therefore, to avoid the noise, we exclude particles that are
 outside of the original radius of the disk.
\section{Results}
We first study isolated galaxy models with different values of
 $\lambda$. We consider four set of simulations followed up to time
 $t=8.0$.
 The parameters and results of runs are presented in the table,
 where $E_0$ and $\bar{E}$ are the initial and mean total energies,
 and
 $\Delta E/E_0$ is the relative change of the total energy during
 the evolution with respect to its initial value.
Though the scale of interaction $\lambda$ and magnitude of
$\alpha$ are unknown, we choose
 their values arbitrarily, such that $\lambda$ is equal to the
 cutoff radius of the disk, bulge and halo for models A1, A2 and A3,
 respectively, and a fixed amplitude of the SF, $\alpha=1$.
 The larger $\lambda$ makes weaker the contribution of
 SF for a fixed galactic size. For $\lambda=\infty$ one obtains the
  Newtonian case, model A4. Previous studies of protogalactic
 interactions under influence of this SF\cite{RodriguezMeza:2009sh}
 were made for scales less than those considered here.

\bigskip
\vspace{0.05in} \noindent \begin{tabular}{|c|c|c|c|c|c|c|} \hline
& & & & & & \\[-0.1in]
 Run & $\lambda$ & $\bar Q$ &$|E_0|$ & $|\bar{E}|$
 & $\mid \frac{\Delta E}{E_0}\mid $,\ \% & $\bar{\eta}$ \\
& & & & & & \\[-0.1in]
\hline
 A1 & 0.4 & 1.0 & 0.7279 & 0.7277 & 0.066 & 0.030 \\
 A2 & 1.0 & 0.9 & 0.9511 & 0.9509 & 0.048 & 0.033 \\
 A3 & 6.0 & 0.9 & 1.1632 & 1.1630 & 0.039 & 0.042 \\
 A4 & $\infty$ & 1.0 & 1.2234 & 1.2232 & 0.036 & 0.045 \\
\hline
\end{tabular}
\smallskip\\
\bigskip

\noindent
All models show a good energy conservation (see table). The
presence of the SF decreases the total potential energy due to
shallower potential well at distances $r>\lambda$.  The
initial models reaccommodate rapidly due to potential
 modification, i.e., shifts to a new equilibrium state. At
the end of evolution, the components of galaxy models with the SF became slightly more
 extended. The velocity profiles of the components match the Newtonian ones
 up to $r\approx\lambda$. For $r>\lambda$
 there is a slow decay in velocities, since the effective
 gravitational constant decreases.
 The distortion parameter
 shows a nearly equal noise level of surface density of the disks.
 The evolution of the Toomre's local stability parameter $Q$ shows
 a slow decay from $Q\approx 1$ to $Q\approx0.9$ with
 mean values presented in the table for each run.

Then, we proceed to study the interaction of two equal galaxy models.
 During the orbital decay we analyze the bar's strength for different
 values of $\lambda$.
 Because the equilibrium galaxy models were constructed with Newtonian
 potential, we  relax them up to time $t=1.0$ with modified SF potential
 in order to reach a new equilibrium state for a given $\lambda$.
 Then we place relaxed galaxies on parabolic orbits and let them interact.

The results are as follows. The first encounter occurs at time
 $t\approx 3.0$. This is the time of major transfer of orbital angular
 momentum, and then disk 1 forms a strong bar,
 while disk 2 is in retrograde motion and develops a very weak bar.
 Figure 1 (left) presents the separation between the centers of mass, $p$,  of two
 galaxies as a function of time. All runs, except A1, after several close
 approaches merge and form a remnant. In the run with model A1, after the first
 approach the galaxies separate such a large distance that they will never encounter
 again. This is because the weaker gravity is, the more lessable the galaxies are to
 become
 bound: at distances larger than $\lambda$ the potential diminishes in comparison
 with the pure Newtonian. In the runs A2 and A3 the SF causes just a retardation
 of the subsequent interactions.  The simulation with model A3 is practically
 identical to Newtonian one.

 In order to analyze the bar formation, we consider following only
 the disk 1.
 In figure 1 (right) we plot the distortion parameters $\eta$ as a function of time
 for runs with galaxy models A2 and A4 only. As a consequence of gravity modification,
 the galaxies do not approach each other too closely as happens in the Newtonian run.
 Thus, weaker perturbations make the bar shorter. The run with galaxy model A3 is very similar
  to the Newtonian case. The bar's phases are displaced
 due to orbit modification.

\begin{widetext}
\begin{figure}
\begin{minipage}{6.5in}
\begin{center}
\includegraphics[width=3.2in]{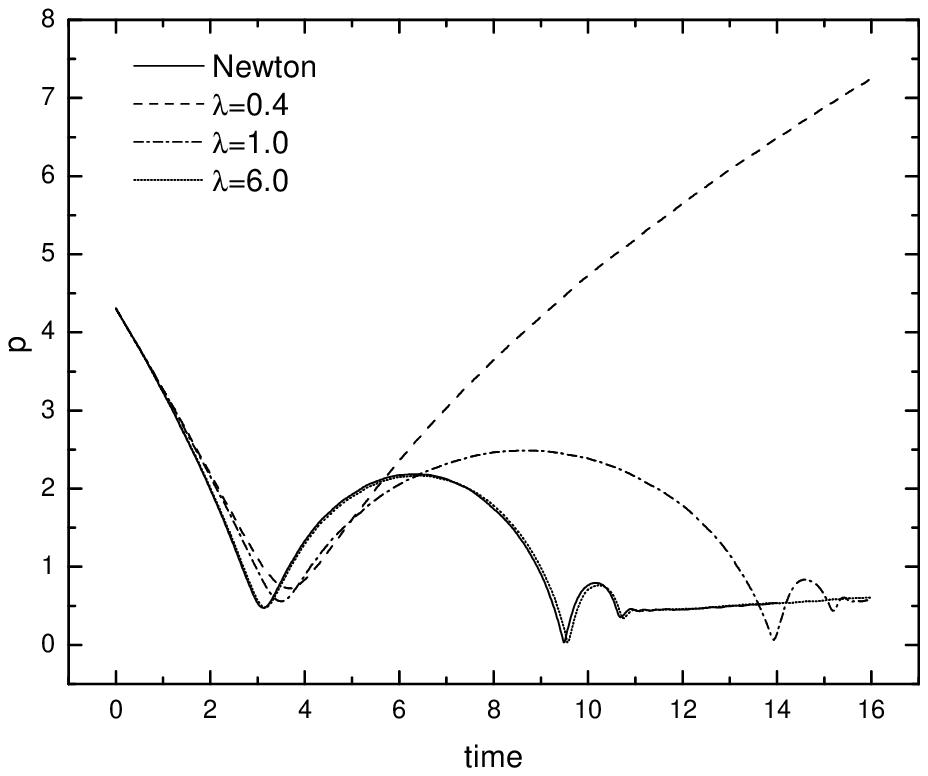}
\includegraphics[width=3.2in]{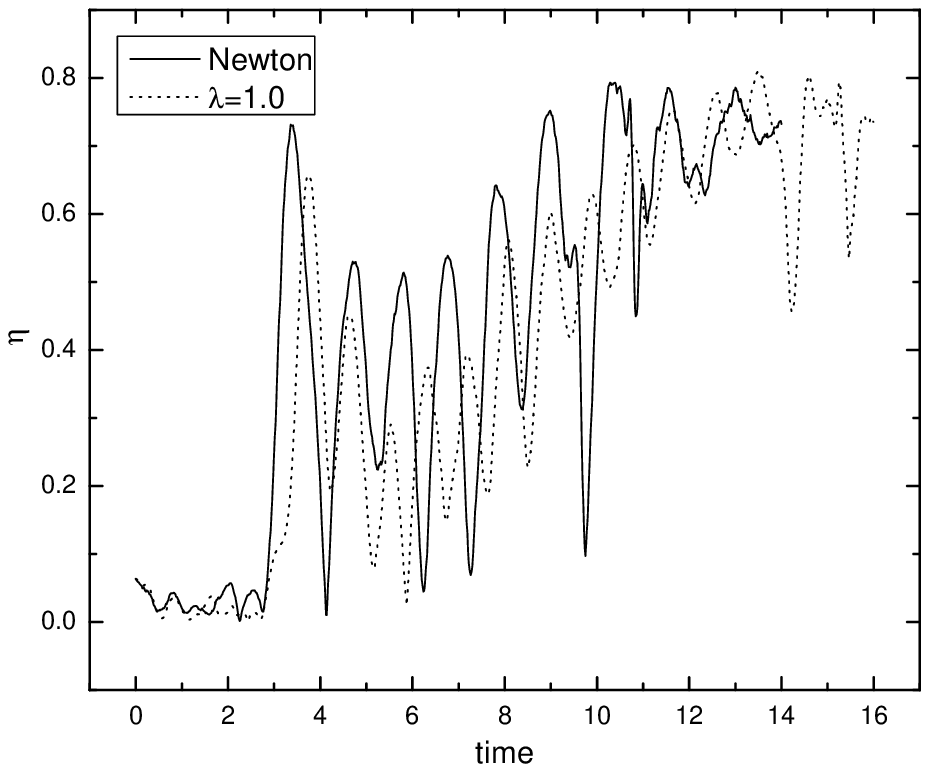}
\vspace{-0.3in}
\end{center}
\caption{Left: The separation between center of mass of two galaxies as a
function of time. Right: Evolution of $\eta$ for two collision
runs with models A1 and A3.}
\end{minipage}
\end{figure}
\end{widetext}


\section{Conclusions}
{}From simulations of isolated galaxy models with different $\lambda$,
 we can see that the addition of a non--minimally coupled SF slightly modifies the equilibrium of Newtonian model,
 acting as a small perturbation, and it diminishes the total potential energy for $r>\lambda$,
 since the effective gravitational constant decreases in this range.
 Our results show that
 the interaction of galaxies with the SF is weaker in comparison with the Newtonian
 case. We have found that the inclusion of the SF changes the dynamical properties
 such as the collision time, bar morphology, and in general the remnant properties.
 All these changes  depend on the pair ($\alpha$, $\lambda$), which
 on the other hand, can be constrained from observations. For
 instance, the duration of interaction
 cannot be larger than the age of the Universe, implying constraints
 on values of $G_0$, which depends on $G$ and $\alpha$.
 These constraints can be provided from
 statistical data on the fraction of observed interacting galaxies.
 A wide range of parameters should be investigated and higher resolution have
 to be used in simulations
 in order to make predictions for particular interacting models.
 Further investigations with more particles and self-consistent initial models are under way.

\section{Acknowledgments}
This work was partly supported by SRE and CONACyT of M\'exico under
contracts U43534-R, SEP-2003-C02-44917.

\end{document}